\documentclass[aps,showpacs,twocolumn,superscriptaddress]{revtex4}
\usepackage{graphics,color}
\usepackage{epsfig}
\usepackage{dcolumn}
\usepackage{bm}
\usepackage{epsfig}
\usepackage{mathrsfs}
\usepackage{amsfonts}
\usepackage{graphicx}
\usepackage{amsmath}
\usepackage{amssymb}
\usepackage{subfigure}
\begin{document}

\title{Many-body Localization Transition of Ising Spin-1 Chains}

\author{Taotao Hu }
\email[email:] {hutt262@nenu.edu.cn}

\author{Yining Zhang }
\affiliation{School of Physics, Northeast Normal University,
	Changchun 130024, People's Republic of China}
\affiliation{School of Physics, Northeast Normal University,
Changchun 130024, People's Republic of China}
\author{Hang Ren}
\affiliation{Key Laboratory of Airborne Optical Imaging and Measurement, Changchun Institute of Optics, Fine Mechanics and Physics, Chinese Academy of Sciences, Changchun 130033,
	People's Republic of China} 
\author{Yiwen Gao }
\affiliation{School of Physics, Northeast Normal University,
	Changchun 130024, People's Republic of China}
\author{Xiaodan Li}
\affiliation{College of Science, University of Shanghai for Science and Technology, Shanghai 200093,  People's Republic of China}
\author{Jiameng Hong }
\affiliation{School of Physics, Northeast Normal University,
	Changchun 130024, People's Republic of China}
\author{Yuting Li}
\affiliation{School of Physics, Northeast Normal University,
	Changchun 130024, People's Republic of China}

\begin{abstract}

  In this paper, we theoretically investigate the many-body localization properties of one-dimensional Ising spin-1 chains by using the methods of exact matrix diagonalization. We compare it with the MBL properties of the Ising spin-1/2 chains. The results indicate that the one-dimensional Ising spin-1 chains can also undergo MBL phase transition. There are various forms of disorder, and we compare the effects of different forms of quasi-disorder and random disorder on many-body localization in this paper. First, we calculate the exctied-state fidelity to study the MBL phase transtion. By changing the form of the quasi-disorder, we study the MBL transition of the system with different forms of quasi-disorder and compare them with those of the random disordered system. The results show that both random disorder and quasi-disorder can cause the MBL phase transition in the one-dimensional Ising spin-1 chains. In order to study the effect of spin interactions, we compare Ising spin-1 chains and spin-1/2 chains with the next-nearest-neighbour(N-N) two-body interactions and the next-next-nearest-neighbour (N-N-N)interactions. The results show that the critical point increases with the addition of the interaction. Then we study the dynamical properties of the model by the dynamical behavior of diagonal entropy (DE), local magnetization and the time evolution of fidelity to further prove the occurrence of MBL phase transition in the disordered Ising spin-1 chains with the (N-N) coupling term and distinguish the ergodic phase (thermal phase) and the many-body localized phase. Lastly, we delve into the impact of periodic driving on one-dimensional Ising spin-1 chains. And we compare it with the results obtained from the Ising spin-1/2 chains. It shows that periodic driving can cause Ising spin-1 chains and Ising spin-1/2 chains to occur the MBL transition.

\end{abstract}

 \maketitle

 \section{Introduction}

 For a long time, physicists have been interested in the dynamical properties of disordered systems. Arbitrarily small disorder induces Anderson localization in one-and two-dimensional systems, as demonstrated in a seminal paper by Anderson in 1958\cite{ref1}. He also pointed out that closed systems for which particles have interactions are similarly unable to thermalize or reach thermal equilibrium when there is sufficiently strong disorder, but instead localize. This theory was demonstrated in a seminal 2006 article\cite{ref2}. This confirms the existence of many-body localization (MBL). A many-body localized system is a perfect model of ``insulation"at non-zero temperatures, which is not subject to thermalization. In summary, the MBL is considered to be a robust counterexample to thermalization, and theoretically, MBL exhibit a new robust accretionability that can be explained in terms of local motion integrals LIOMs\cite{ref3}. A wide set of localization motion integrals emerges, which provides new ideas for the study of many-body localization\cite{ref4,ref5}. Since many-body localization preserves coherence and certain forms of topological order, it is more conducive to quantum computation and has a wide range of research value and application prospects.

 At present, the MBL phase transition has attracted much attention\cite{ref6,ref7,ref8,ref9,ref10,ref11,ref12,ref13,ref14,ref15,ref16,ref17,ref18,ref19,ref20}. Inspired by Anderson’s localization caused by the disorder, people study many-body localization caused by uniform disorder\cite{ref21,ref22} and periodic disorder\cite{ref23}. For example disorder leads to MBL phase transitions in Ising spin-1/2 chains systems. It is shown that if the random disorder field in the system is weak, the eigenstates of the system are thermal and the system is in the ergodic phase; if the random disorder field in the system is strong, the eigenstates of the system are localized and the system is in the localized phase\cite{ref29}. It is shown that a quasi-periodic potential system can also have the MBL phase\cite{ref24}, and the characteristics of the phase were observed in cold atom experiments\cite{ref25,ref26,ref27,ref28}. By providing a detailed finite-size scaling analysis of the Quasi-Periodic Potential (QP)-MBL transiation, along with a comparison to the random MBL transiation. The MBL of the quasi-periodic potential model is found to be more stable compared to the random model, which is opposite to the trend for single-particle localization.  
 
 Many characteristic dynamical properties of MBL have been observed in various studies. Some of these properties include: the power-law decay of imbalance\cite{ref30,ref31,ref32,ref33}, quantum Fisher information\cite{ref34,ref35,ref36}, local magnetization\cite{ref37}, and the logarithmic spreading of entanglement entropy (EE)\cite{ref38,ref39,ref40,ref41,ref42}. However measuring EE in large quantum systems, especially those involving many qubits, can be quite challenging. Because the measurement of EE requires quantum state tomography\cite{ref43}. People have found alternative probes to MBL other than EE. Recently It has been shown that the dynamical behavior of diagonal entropy (DE) also exhibits slow growth in MBL systems \cite{ref44}. Even more attractive is the fact that only the diagonal elements of the density matrix are needed to calculate the DE. In the MBL states, the information of the initial state can be well protected\cite{ref45,ref46}. Therefore, MBL has become a popular direction in the field of quantum many-body physics. 
 
 Similar to the effect of the disordered field on the phase of a closed system, people argue that a  periodic driving can likewise drive a closed system so that it undergoes a phase transition between the localized phase and the ergodic phase. The long-time behavior of the periodically driven system, i.e., the time evolution of the system, is determined by the properties of the so-called Floquet Hamiltonian $\hat{H}_{F}$, defined according to the Floquet operator\cite{ref47,ref48}, which is independent of time,
 \begin{equation}
	\hat{F}=e^{-i \hat{H}_{F} T}
\end{equation}
This is the evolution operator over a period of time, 
  \begin{equation}
 	\hat{F}=\mathcal{T} \exp \left\{-i \int_{0}^{T} H(t) d t\right\}
 \end{equation}
The Hamiltonian of a periodically driven system is a periodic function of time, $H(t+T)=H(t)$. Here $\mathcal{T} \exp$ is a time-ordered exponential.

 In condensed matter physics, spin-1 and spin-1/2 systems refer to different models of quantum spin. The ground state of the spin-1 chains can be understood as the ground state of each spin chains. Spin-1 is considered as a symmetric combination of two spin-1/2\cite{ref49,ref50}. While spin-1/2 chains were often studied in the past, in recent years spin-1 chains have been frequently used as theoretical models for studying various phenomena, including magnetic and quantum phase transitions\cite{ref51,ref52,ref53,ref54,ref55,ref56}. In 2023\cite{ref57}, Wajid Joyia and Khalid Khan et al. studied spin-1/2 and spin-1 systems which eventually exhibit quantum critical behaviour with different numbers of iterations. The quantum critical behaviour in spin-1/2 and spin-1 systems was further observed through the unresolved and scalar behaviour of quantum coherence. The results show that spin-1 system reaches the quantum critical point with fewer Quantum re-normalization group method compared to the spin-1/2 system. 
 
 However, there are few studies on the many-body localization properties of the Ising spin-1 chains. Do spin properties affect the MBL phase transition? Therefore, this paper focuses on the MBL phase transition of the Ising spin-1 chains. We first study properties of the eigenstates of the one-dimensional Ising spin-1 chains through the excited-state fidelity, and further investigate the properties of the eigenstates of the one-dimensional Ising model by varying the disordered form and incorporating interactions. In addition, we also compare the MBL properties of spin-1 chains with those of spin-1/2 chains. Then we study the dynamical properties of the disordered Ising spin-1 chains with the (N-N) coupling term by the dynamical behavior of diagonal entropy (DE), local magnetization and the time evolution of fidelity to further prove the occurrence of MBL phase transition in the model and distinguish the ergodic phase (thermal phase) and the many-body localized phase. Finally we also discuss the effect of periodic driving on the one-dimensional Ising spin-1 chains. We also compare the MBL properties of spin-1 chains with those of spin-1/2 chains.

\section{ Model used for numerics.}

We study a specific simple spin model, a one-dimensional Ising model with disordered fields along the $z$ direction. Considering open boundary conditions. The Hamiltonian can be written as

\begin{equation}
	H_{0}= \sum_{i=1}^{N-1} {S_{i}}^{z}{S}_{i+1}^{z}+\sum_{i=1}^{N} h_{i} S_{i}^{z}
\end{equation}
where $h_i$ is a disordered variable at lattice point $i$. In this paper, we will discuss random disorder and two forms of quasi-disorder. In random disorder, $h_i$ is the random disordered variable at position $i$ and each $h_i$ has a probability distributed between $[-h,h]$. 
For the quasi-disordered,
\begin{equation}
	h_i= h\cos (\frac{{{\pi}\sqrt{2}}}{2}N+\phi)
\end{equation}
there are two ways to value $\phi$. One is that $\phi$ is a random phase taken from the uniform distribution between [$-\pi$, $\pi$]. And the other is that we choose these points as $\phi_i$ and increase the first value $\phi_1$ by a same small amount each time the disorder is realized, then the quasi-disordered field of the many-body system is realized.

\section{ Results and discussion } 
Fidelity is an important concept in quantum information theory \cite{ref58} and quantum chaos \cite{ref59}. It is used to denote the extent to which the initial state is maintained during information transmission, and it is also a very important parameter to measure the quality of communication. It also plays an important role in the study of quantum phase transitions. In this paper we use excited state fidelity to study the many-body localized phase transition. The fidelity of the n-th excited state $\psi_{n}(h)$ is defined as the overlap of the excited states with parameters $h$ and $h$ + $\delta h$, here $\delta h$ is a small shift of this field\cite{ref60}.

\begin{equation}\
  F_{n}(h,h+\delta h)=\left|\left\langle\psi_{n}(h)\mid\psi_{n}(h+\delta h)\right\rangle\right|
\end{equation}

To test the fidelity of excited states, for the small parameter perturbation $\delta h_{i}$ for each site, we let $\delta h_{i}$ = $\varepsilon h_{i}$ ($\varepsilon=10^{-3}$). It is worth noting that the parameter perturbation $\delta h_{i}$ for each site are also different random variables.
In general, the fidelity value is close to 1, but near the critical point of phase transition, there will be significant changes in the fidelity value.

For each disorder realization, we find almost all the many-body eigenstates $|\psi_{n}\rangle$ \cite{ref61}. We then compute the fidelity $F_{n}$ for each eigenstate $|\psi_{n}\rangle$. Averaging over all selected excited states and disorder realizations yields the mean value $\mathsf{E}[F]$. For each disorder amplitude $h$, we used 1000 disorder realizations for $N=6$, 500 realizations for $N=8$ in this paper. Then we use exact matrix diagonalization for numerical analysis to obtain the data in this paper. The parallel programming techniques were employed to make computations feasible.

We investigated the effect of three types of disorder on the MBL phase transition of Ising spin-1 with system sizes N=6 and N=8. In Fig.1 and Fig.2, we plot the average excited-state fidelity $ \mathsf{E}[F] $  as a function of the disorder strength $h$ to observe whether the MBL phase transition occurs in the Ising spin-1 chains with different forms of disorder when the system size are N=6 and N=8 respectively. Based on previous studies\cite{ref62,ref63}, we can roughly get the critical point of the MBL phase transition from the inflection points of fidelity curves in excited states of the MBL phase transition. As seen in Fig.1 and Fig.2, $ \mathsf{E}[F] $ is close to 1 at the beginning, in the ergodic phase (small $h$), as the disorder strength $h$ increases, $ \mathsf{E}[F] $ gradually decreases, until $h$ approaches the critical point $h_{c}$. Afterward, in the localized phase (large $h$), as $h$ continues to increase, $ \mathsf{E}[F] $ gradually increases and finally stabilizes. This indicates that the system undergoes a phase transition from the ergodic phase to the localized phase. As seen in Fig.1 and Fig.2, the critical point is about $h_{c}$= 1.3 for the random disordered system when system size is N=6, the critical point is about $h_{c}$= 1.5 for the random disordered system when system size is N=8.

\begin{figure}

\centering
\includegraphics[width=0.50\textwidth]{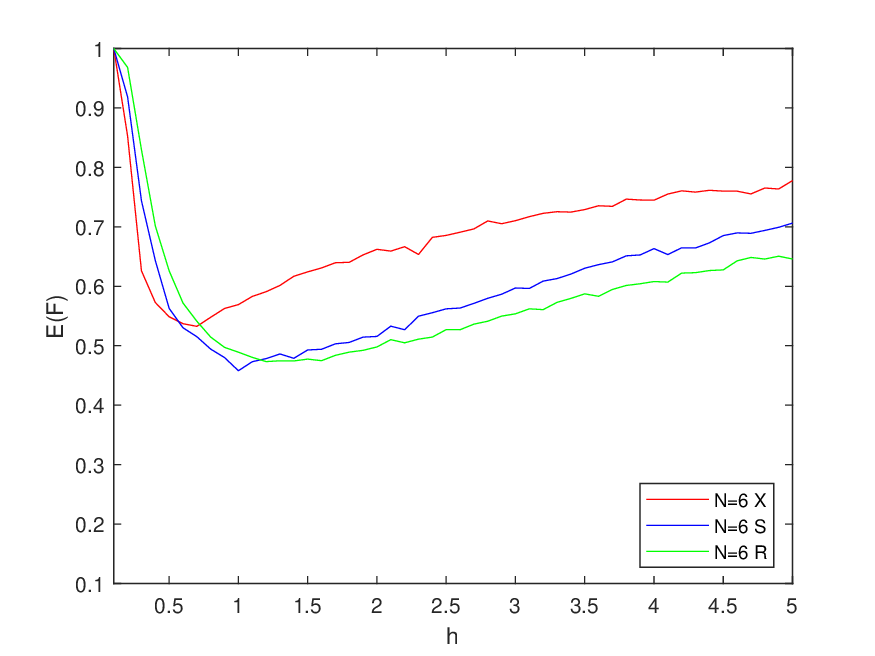}
\caption{The averaged fidelity  $ \mathsf{E}[F] $  as a function of the disorder strength $h$ for different forms of disorder in the Ising spin-1 chains. The system size $N$ are indicated in the legend. The red line represents the quasi-disorder form that $\phi$ increases with a small amount for each disorder realization, the blue line represents the quasi-disorder form that $\phi$ is a random number, the green line represents random disorder. }
\label{fig1 }

\end{figure}

\begin{figure}

\centering
\includegraphics[width=0.50\textwidth]{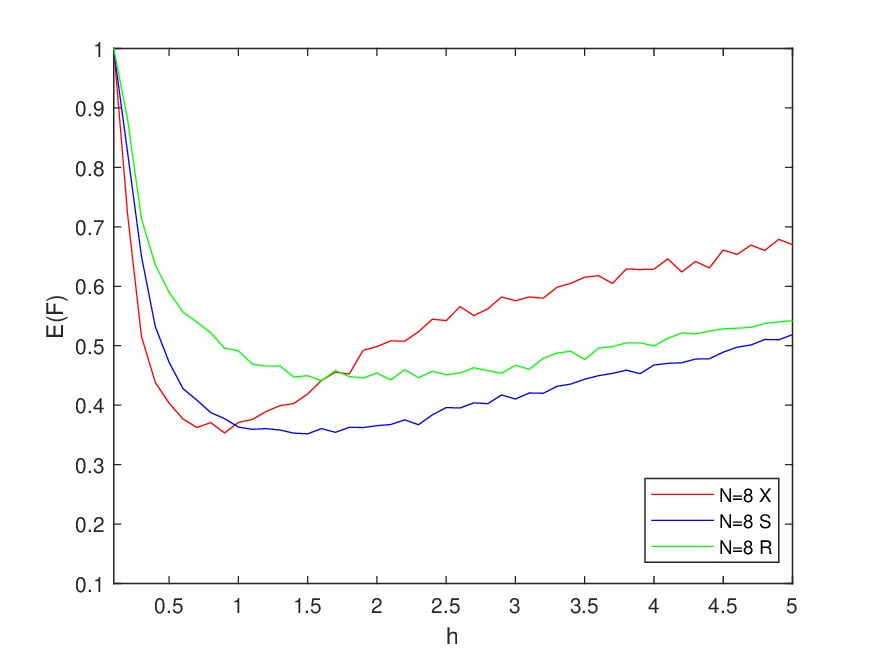}
\caption{The averaged fidelity  $ \mathsf{E}[F] $  as a function of the disorder strength $h$ for different forms of disorder in the Ising spin-1 chains. The system size $N$ are indicated in the legend. The red line represents the quasi-disorder form that $\phi$ increases with a small amount for each disorder realization, the blue line represents the quasi-disorder form that $\phi$ is a random number, the green line represents random disorder. } 
\label{fig2 }

\end{figure}

It can be seen from Fig.1 and Fig.2 that both quasi-disorder and random disorder can cause the MBL phase transition in the Ising spin-1 chains. However, the critical points of the MBL phase transition are different for system with different types of disorder, and for Ising spin-1 chains with different system size, the critical points are also different. Furthermore, to investigate the influence of spin properties on the phase transition, we conducted a comparative study on the many-body localization properties of Ising spin-1/2 chains with three types of disorder which is shown in Fig.3 and Fig.4. They all show that, for the quasi-disordered spin-1/2 model, the curves of the excited-state fidelity fluctuate significantly with the variation of the disorder strength $h$. However, they still indicate the transition of the system from the ergodic phase to the localized phase. These results are similar to the MBL phase transition in the Ising spin-1 chains with three types of disorder shown in Fig.1 and Fig.2.

\begin{figure}

\centering
\includegraphics[width=0.50\textwidth]{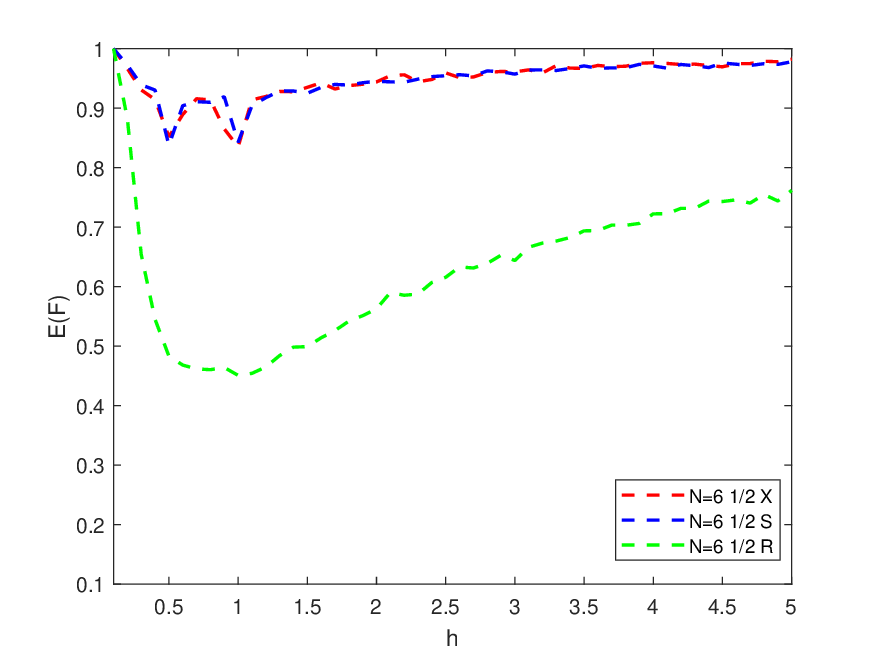}
\caption{The averaged fidelity  $ \mathsf{E}[F] $  as a function of the disorder strength $h$ for different forms of disorder in the Ising spin-1/2 chains. The system size $N$ are indicated in the legend. The red line represents the quasi-disorder form that $\phi$ increases with a small amount for each disorder realization, the blue line represents the quasi-disorder form that $\phi$ is a random number, the green line represents random disorder. } 
\label{fig3 }

\end{figure}

\begin{figure}

\centering
\includegraphics[width=0.50\textwidth]{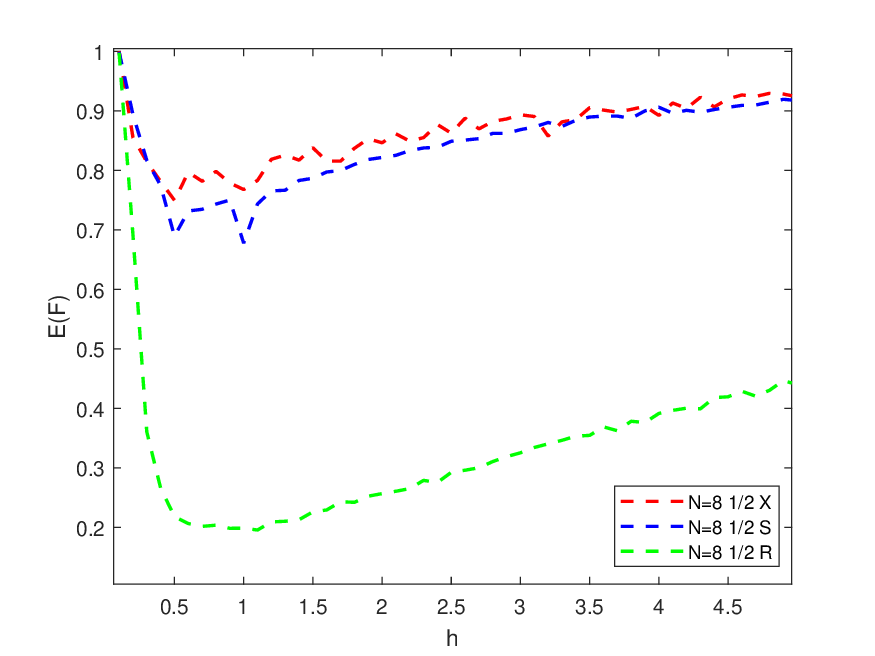}
\caption{The averaged fidelity  $ \mathsf{E}[F] $  as a function of the disorder strength $h$ for different forms of disorder in the Ising spin-1/2 chains. The system size $N$ are indicated in the legend. The red line represents the quasi-disorder form that $\phi$ increases with a small amount for each disorder realization, the blue line represents the quasi-disorder form that $\phi$ is a random number, the green line represents random disorder. } 
\label{fig4 }

\end{figure}

Considering the effect of the spin properties on the MBL phase transition, we compared and studied the MBL phase transitions of the two types of spins when system sizes are 6 and 8 in Fig.5 and Fig.6 respectively. They show that under the same conditions, the critical points of MBL phase transition in the spin-1 chains with three types of disorder are different from those in the spin-1/2 chains.
\begin{figure}

\centering
\includegraphics[width=0.50\textwidth]{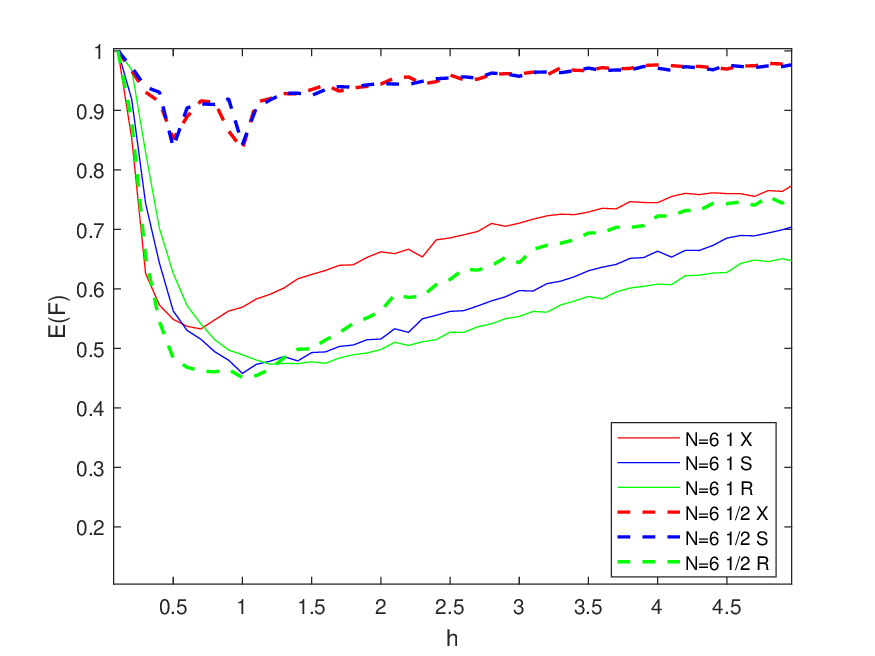}
\caption{The averaged fidelity $ \mathsf{E}[F] $  as a function of the disorder strength $h$ for different forms of disorder in two spin chains. The system size is N=6.  The red line represents the quasi-disorder form that $\phi$ increases with a small amount for each disorder realization, the blue line represents the quasi-disorder form that $\phi$ is a random number, the green line represents random disorder. Solid lines represent spin-1 and dotted lines represent spin-1/2.} 
\label{FIG5 }

\end{figure}
\begin{figure}

\centering
\includegraphics[width=0.50\textwidth]{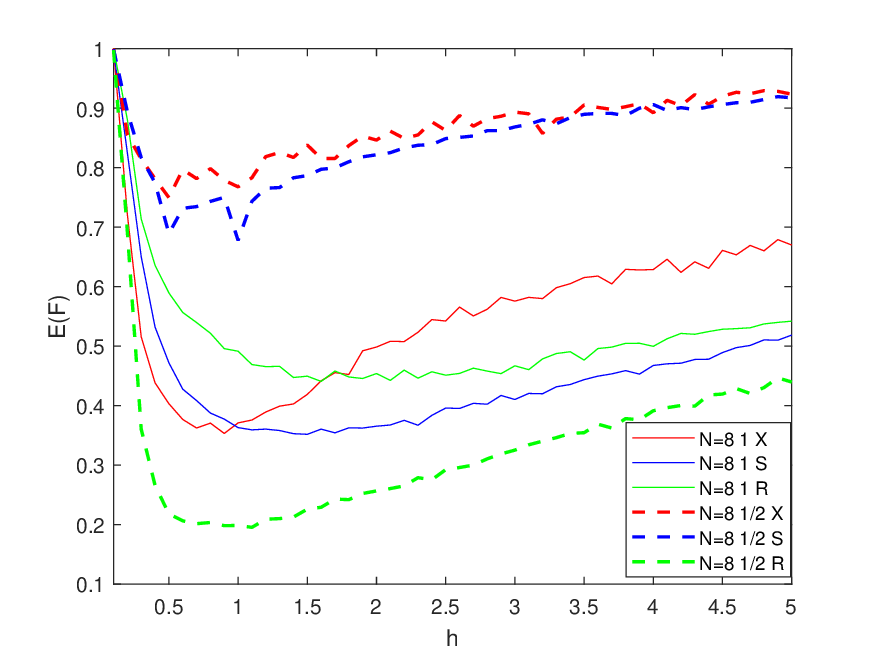}
\caption{The averaged fidelity $ \mathsf{E}[F] $ as a function of the disorder strength $h$ for different forms of disorder in two spin chains. The system size is N=8. The red line represents the quasi-disorder form that $\phi$ increases with a small amount for each disorder realization, the blue line represents the quasi-disorder form that $\phi$ is a random number, the green line represents random disorder. Solid lines represent spin-1 and dotted lines represent spin-1/2. } 
\label{fig6 }

\end{figure}
\begin{figure}

\centering
\includegraphics[width=0.50\textwidth]{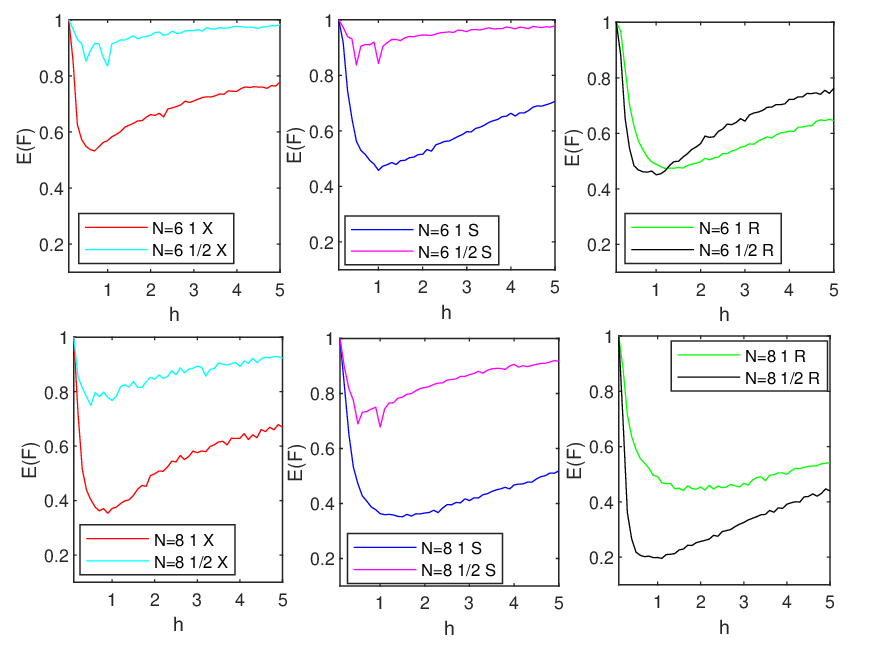}
\caption{The averaged fidelity $ \mathsf{E}[F] $ as a function of disorder strength $h$ of two spin chains with the same disorder. The system size of the above three images is N=6, while the system size of the below three images is N=8. Spin-1 and spin-1/2 are labeled in the illustration.} 
\label{fig7 }

\end{figure}

In order to facilitate the observation and investigate the special properties of the Ising model, we plot Fig.7 to study the MBL phase transitions of the spin-1 and spin-1/2 chains with the same disorder. In Fig.7, we plot the fidelity $ \mathsf{E}[F] $ of the different spin chains as a function of the disorder strength $h$ for three different types of disorder. The three images above are for the system size is N=6, and the following three images are for the system size is N=8. The results show that the critical points of MBL phase transition in the spin-1 chains are larger than those in the spin-1/2 chains for both quasi-disordered system and random disordered system. It means thay the MBL phase transition of the spin-1 chains requires a larger degree of disorder.
By comparing the top and bottom plots in Fig.7, it can be seen that the MBL phase transition critical point increases with the increase of the system size N for both the spin-1 chains and the spin-1/2 chains.

Here we consider only the nearest-neighbor interactions, and in order to further investigate the effect of spin interactions of the disorder Ising model, we add the next-nearest-neighbor (N-N)interactions, and the next-next-nearest-neighbor (N-N-N) interactions. The Hamiltonian can be written as
\begin{equation}
	H_{0}= \sum_{i=1}^{N-2} {S_{i}}^{x}{S}_{i+2}^{x}+{S_{i}}^{y}{S}_{i+2}^{y}+\sum_{i=1}^{N-1}{S_{i}}^{z} {S}_{i+1}^{z}+\sum_{i=1}^{N} h_{i} S_{i}^{z}
\end{equation}
and
\begin{align}
&H_{0}= \sum_{i=1}^{N-3}{S_{i}}^{x}{S}_{i+3}^{x}+{S_{i}}^{y}{S}_{i+3}^{y} \nonumber\\
&+\sum_{i=1}^{N-2} {S_{i}}^{x}{S}_{i+2}^{x}+{S_{i}}^{y}{S}_{i+2}^{y} +\sum_{i=1}^{N-1}{S_{i}}^{z} {S}_{i+1}^{z}
 +\sum_{i=1}^{N} h_{i} S_{i}^{z}
\end{align}

We plot Fig.8 and Fig.9, they show that the critical disorder strength of the phase transition increase for both the spin-1 and spin-1/2 chains with the addition of the interaction. This means that as the interaction adding, the system needs stronger disorder, the critical point of the MBL phase transition will be increased. 

\begin{figure}

\centering
\includegraphics[width=0.50\textwidth]{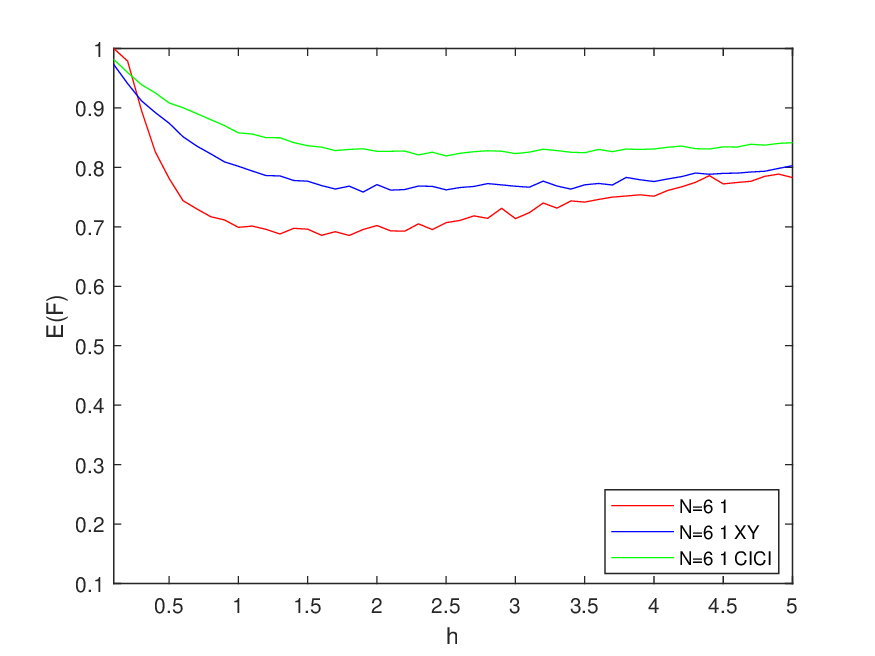}
\caption{The averaged fidelity $ \mathsf{E}[F] $ as a function of disorder strength $h$ of random disorder in the Ising spin-1 chains, The system size is N=6. The red curve represents the the nearest-neighbor interactions, the blue curve represents the (N-N) interactions, the green curve represents the (N-N-N) interactions. } 
\label{fig8 }

\end{figure}
\begin{figure}

\centering
\includegraphics[width=0.50\textwidth]{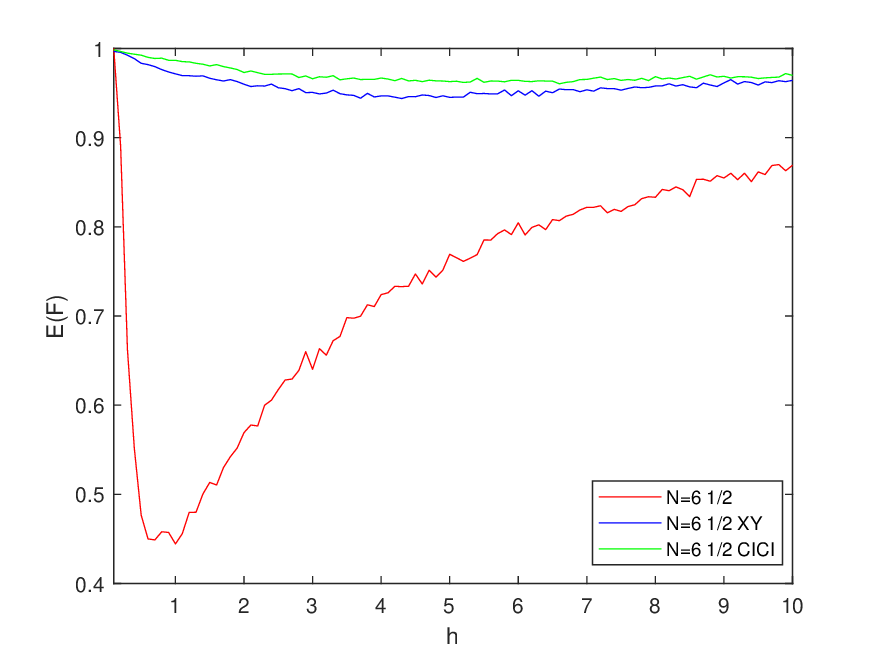}
\caption{The averaged fidelity $ \mathsf{E}[F] $ as a function of disorder strength $h$ of random disorder in the Ising spin-1/2 chains, The system size is N=6. The red curve represents the the nearest-neighbor interactions, the blue curve represents the (N-N)  interactions, the green curve represents the (N-N-N) interactions. } 
\label{fig9 }

\end{figure}

We further investigated the dynamical properties of the disordered Ising spin-1 chain with the (N-N) coupling term by examining the dynamical behavior of diagonal entropy (DE), local magnetization, and the time evolution of fidelity. This additional analysis aims to reinforce the evidence of the occurrence of MBL phase transition in this model, and to distinguish between the ergodic phase (thermal phase) and the many-body localized phase. The system is initially prepared in the $\rm{N\acute{e}er}$ (product) state $|\psi(0)\rangle$ at $t=0$, then one can get the $|\psi(t)\rangle=e^{-iHt}|\psi(0)\rangle$ after evolving $|\psi(0)\rangle$ with $H$ in Eq.(6) for time $t$. We commence by computing the dynamics of diagonal entropy (DE) under the influence of a disordered external field. The DE takes the form\cite{ref44}:
\begin{equation}
{S}^{\rm{diag}}(t)=-{\rm{Tr}}(\rho(t)^{\rm{diag}}{\rm{ln}}\rho(t)^{\rm{diag}}),
\end{equation}
where $\rho(t)^{\rm{diag}}$ is the matrix obtained from the pure state density
operator $\rho(t)=|\psi(t)\rangle\langle\psi(t)|$ by deleting its off-diagonal elements. Averaging over all disorder realizations yields the mean value $E[S^{\rm{diag}}(t)]$, 2000 disorder realizations for $N=6$ is used here.

The long-time dynamics of DE is illustrated in Fig.10. According to Fig.8, we have obtained that the critical points are $h_c=2.2$  for the Ising spin-1 chains with system size $N=6$. Therefore we selected two sets of values, one for $(h<h_c)$ (h=0.1,1) and one for $(h>h_c)$ (h=5,10,15), and plotted the following figures. As depicted in Fig.10, in the ergodic phase $(h<h_c)$, the DE initially undergoes rapid power law growth, followed by saturation to a substantial value within a relatively brief time frame. In contrast, within the localized phase $(h>h_c)$, we observe a gradual logarithmic growth in DE, eventually saturating to a lower value over an extended duration.

\begin{figure}
\centering
\includegraphics[width=0.50\textwidth]{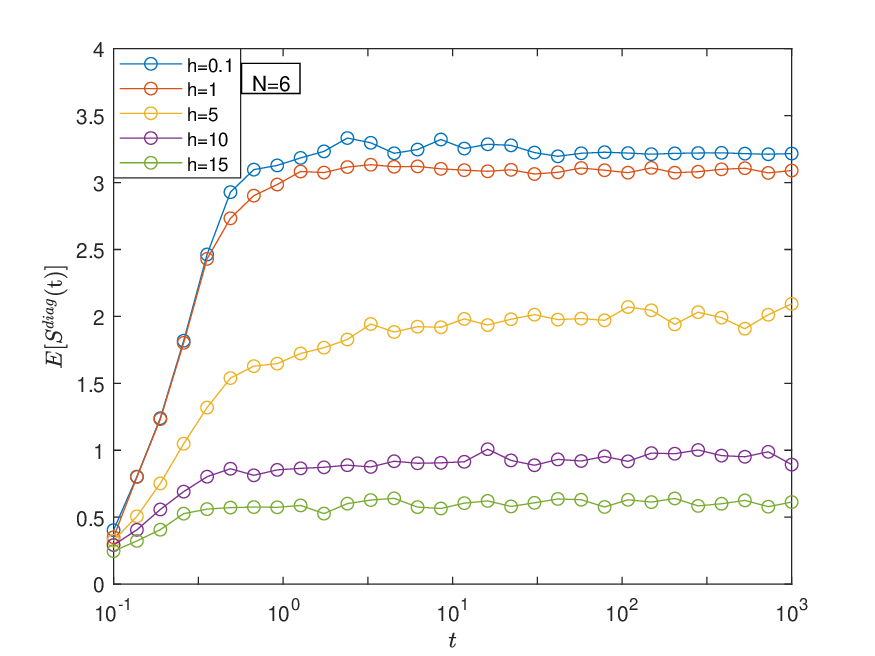}
\caption{Plot of averaged diagonal entropy $E[S^{\rm{diag}}(t)]$ for the random disordered Ising spin-1 chains with $N = 6$ as a function of time $t$. The values of the disorder strength $h$ are indicated in the legend. In the ergodic phase (small $h$) the DE first experiences a fast power-law growth, and then saturate to a large value while it only has slow logarithmic growth in the localized phase (large $h$).} 
\label{fig9 }

\end{figure} 
We continue our investigation into the evolution of local magnetization, computing the expectation value of the spin on a given site ${I}$ at time $t$ to obtain the magnetization ${M_{I}}(t)$, expressed as\cite{ref21}:
\begin{equation}
{M_{I}}(t)=\langle\psi(t)|S_{i}^{z}|\psi(t)\rangle
\end{equation}

Averaging over all disorder realizations yields the mean value $E[{M_{I}}(t)]$, 2000 disorder realizations for $N=6$ is used here. Fig.11 illustrate the time evolution of $E[{M_{I}}(t)]$. It shows that in the ergodic phase $(h<h_c)$, the local magnetization ${M_{I}}(t)$ initially experiences a significant decline in a relatively shorter time, followed by saturation to a small value. In the localized phase $(h>h_c)$, ${M_{I}}(t)$ decays less over a relatively longer period before saturating at a larger value. Moreover, when the disorder strength $h$ is large enough ($h$=10,15), ${M_{I}}(t)$ remains nearly unchanged. What is more, the smaller disorder strength, the larger the decay of ${M_{I}}(t)$, and the larger disorder strength, the less the decline of ${M_{I}}(t)$. 

\begin{figure}
\centering
\includegraphics[width=0.50\textwidth]{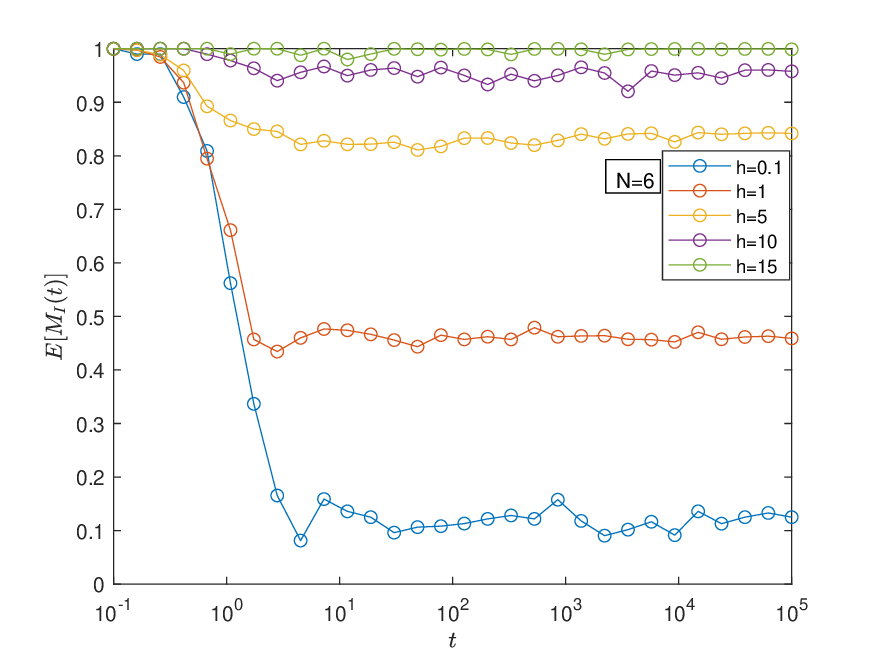}
\caption{Averaged local magnetization $E[{M_{I}}(t)]$ at a given site $I$ for the random disordered Ising spin-1 chains with $N = 6$ as a function of time $t$. The values of the disorder strength $h$ are indicated in the legend. In the ergodic phase (small $h$) the magnetization ${M_{I}}(t)$ first has relatively large decline, and then saturates to a small value while it only has less decline in the localized phasen(large $h$). } 
\label{fig9 }

\end{figure} 

Next we also study the time evolution of Fidelity $F(t)$, for mixed states, it is defined as\cite{ref62}
\begin{equation}
{F}(t)={\rm{Tr}}[\rho(t)^{1/2}\rho(0)\rho(t)^{1/2}]^{1/2},
\end{equation}
where $\rho(t)=|\psi(t)\rangle\langle\psi(t)|$, $\rho(0)=|\psi(0)\rangle\langle\psi(0)|$. Averaging over all disorder realizations yields the mean value $E[F(t)]$, 2000 disorder realizations for $N=6$ is used here. Fig.12 illustrates the time evolution of $E[F(t)]$. In the ergodic phase $(h<h_c)$, the fidelity $F(t)$ initially experiences a rapid decay, stabilizing at a relatively low value after long-time evolution. In the localized phase $(h>h_c)$, we observe the less decline of $F(t)$, and the larger disorder strength, the less the decline of $F(t)$. This means that in the localized phase, the information of the initial state can be well protected if the disorder strength of the system is large enough. Additionally, it also shows that as $h$ increases, the preservation of initial state information becomes higher, in line with the MBL hypothesis.

Through the dynamical behavior of diagonal entropy (DE), local magnetization and the time evolution of fidelity, we can further prove the occurrence of MBL phase transition in the Ising spin-1 chains and distinguish the ergodic phase and the localized phase. And the results also show that the disordered Ising spin-1 chains system can protect its initial information well in the localized phase if the disorder strength of the system is large enough.

\begin{figure}
\centering
\includegraphics[width=0.50\textwidth]{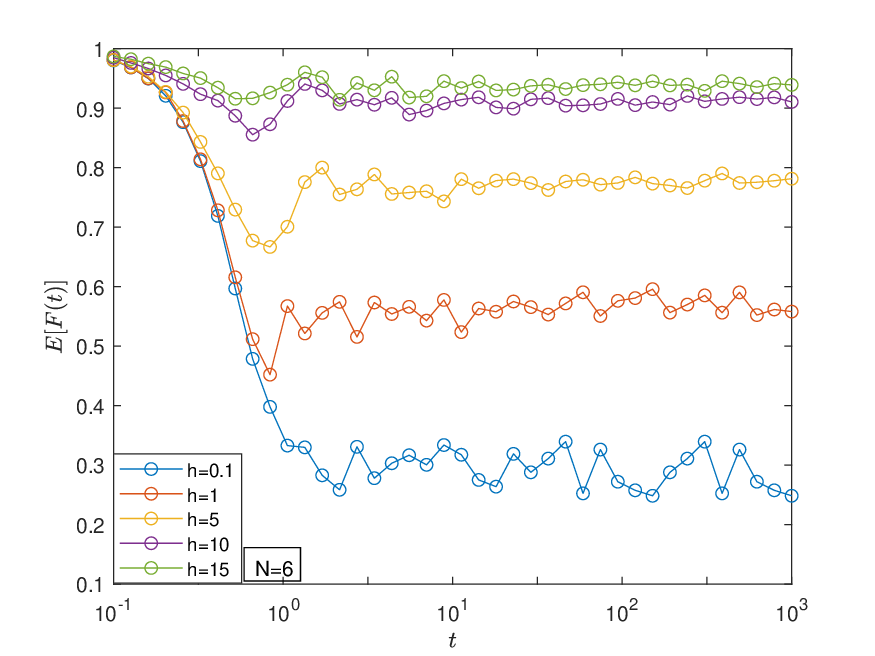}
\caption{The averaged time evolution of Fidelity $E[F(t)]$ for the random disordered Ising spin-1 chains with $N = 6$ as a function of time $t$. The values of the disorder strength $h$ are indicated in the legend. In the ergodic phase (small $h$) the Fidelity $F(t)$ first experiences a fast decay, and then saturates to a relatively low value while it only has less decline in the localized phase (large $h$). } 
\label{fig9 }

\end{figure}

In order to further explore the properties of many-body localization in Ising model, we add periodic driving to the system to study the effect of periodic driving on many-body localization.
We add a time-periodic field in the form of a trigonometric function to the Ising model and obtain the periodically driven system as follows:

\begin{equation}
	H(t)=H_{0}+\cos\omega t \sum_{i=1}^{N}  S_{i}^{z}
\end{equation}

where $H_{0}$ is the Hamiltonian in Eq.(3). By adjusting disorder strength $h$, we can make the system initially in the ergodic phase and the localized phase respectively. From Fig.13 to Fig.16, we investigate the effect of periodic driving on the properties of MBL in these random disordered systems. The system size is $N=6$. We can see that $ \mathsf{E}[F] $ decrease as the driving period $T$ increases, which characterizes the occurrence of the phase transition. 

As seen in Fig.1, the critical point is about $h_c=1.5$ for the random disordered system when system size is $N=6$. In Fig.13, $h>h_c$, the system is initially in the localized phase, it shows that the periodic driving can also cause the disordered Ising spin-1 chains to occur a phase transition from the localized phase to the ergodic phase, but the curves exhibit significant fluctuations, here the critical driving period $T_{c}$\cite{ref65} is smaller for systems with larger $h$. It shows that the system with larger $h$ is closer to the phase transition threshold and it is more easily to occur transition. In Fig.14, $h<h_c$, the system is initially in the ergodic phase, it shows that the periodic driving can also cause the disordered Ising spin-1 chains to occur a phase transition from the ergodic phase to the localized phase, but the curves exhibit significant fluctuations, here the critical driving period $T_{c}$ are smaller for systems with larger $h$. In Fig.14, we can also observe that as the disorder strength $h$ increases, the fluctuation of the curves decreases, indicating a more stable phase transition. In summary, the periodic driving can cause the disordered Ising spin-1 chains to occur a phase transition, and regardless of whether the system initially resides in the ergodic or localized phase, for systems with larger $h$, the critical driving period $T_{c}$ for the transition becomes smaller, implying it is more easily to occur transition. 

We also compare the MBL properties of periodically driven spin-1 chains with those of spin-1/2 chains. In Fig.15 and Fig.16, we plot the averaged excited state fidelity $ \mathsf{E}[F] $ as a function of driving period $T$ in the Ising spin-1/2 chains. It shows that whether the system is initially in the ergodic phase or the localized phase, the system with larger $h$ is closer to the phase transition threshold and it is more easily to occur transition. The results are consistent with Ising spin-1 chains. However, the critical driving period $T_c$ corresponding to different spin properties is also different. Comparing Fig. 14 and Fig. 16, it shows that the spin-1/2 chains can be driven more stably from the ergodic phase to the localized phase.

\begin{figure}

\centering
\includegraphics[width=0.50\textwidth]{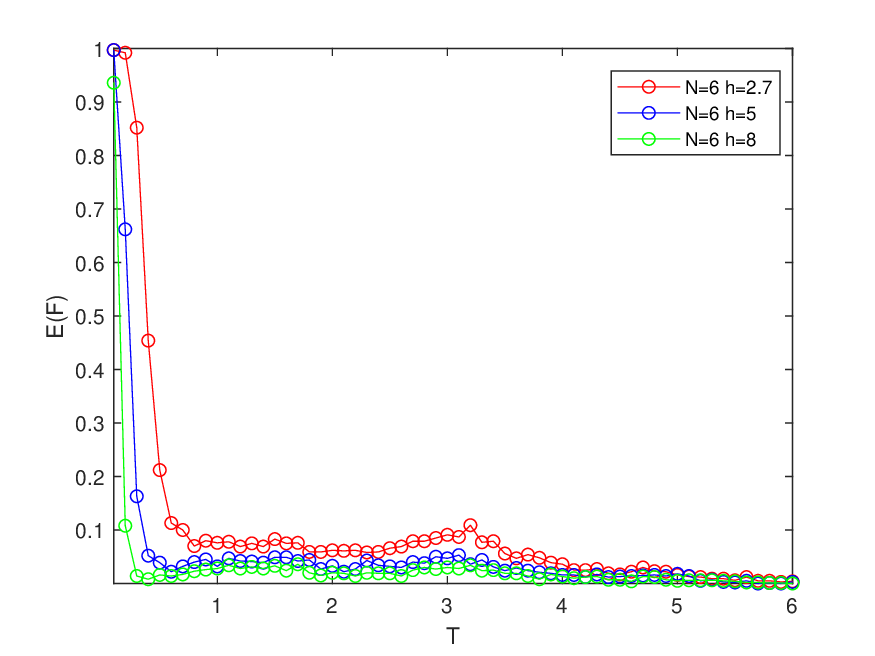}
\caption{The averaged fidelity $ \mathsf{E}[F] $ as a function of driving period $T$ of random disorder in the Ising spin-1 chains. The system size is N=6. The quasi-disorder strength $h$ are indicated in the legend. } 
\label{fig9 }

\end{figure}
\begin{figure}

\centering
\includegraphics[width=0.50\textwidth]{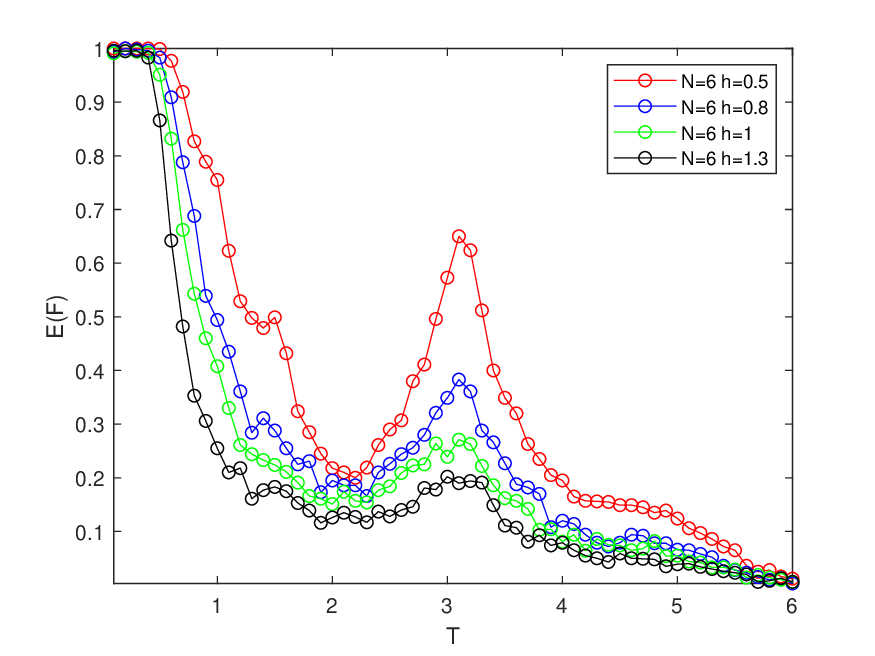}
\caption{The averaged fidelity $ \mathsf{E}[F] $ as a function of driving period $T$ of random disorder in the Ising spin-1 chains, The system size is N=6. The quasi-disorder strength $h$ are indicated in the legend. } 
\label{fig9 }

\end{figure} 
\begin{figure}

\centering
\includegraphics[width=0.50\textwidth]{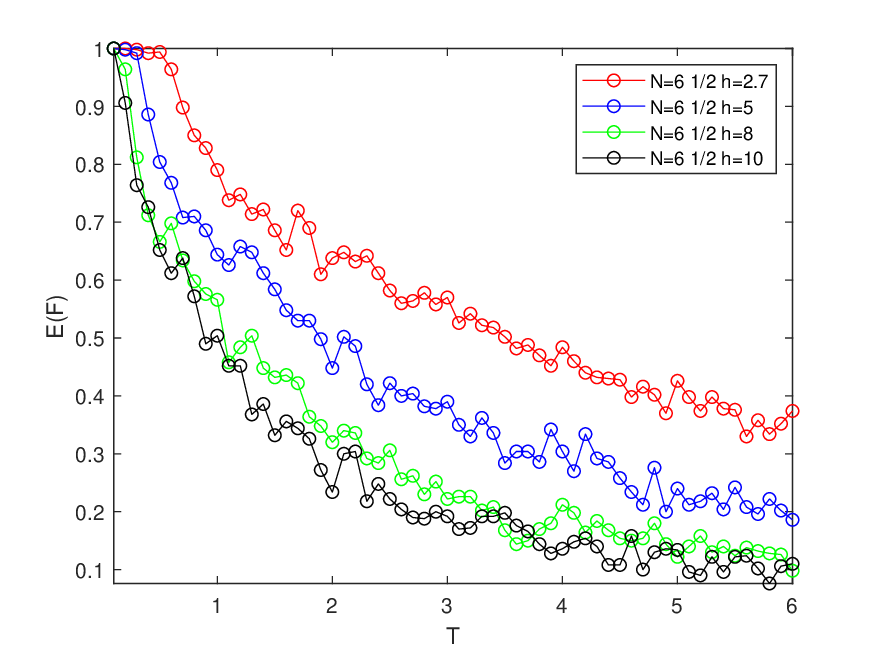}
\caption{The averaged fidelity $ \mathsf{E}[F] $ as a function of driving period $T$ of random disorder in the Ising spin-1/2 chains, The system size is N=6. The quasi-disorder strength $h$ are indicated in the legend. } 
\label{fig9 }

\end{figure} 

\begin{figure}

\centering
\includegraphics[width=0.50\textwidth]{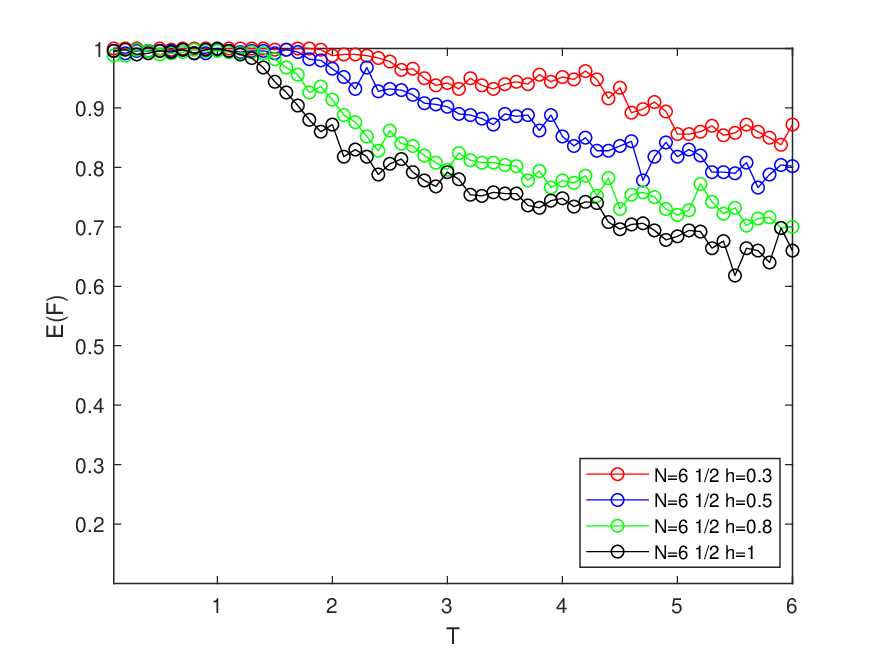}
\caption{The averaged fidelity $ \mathsf{E}[F] $ as a function of driving period $T$ of random disorder in the Ising spin-1/2 chains, The system size is N=6. The quasi-disorder strength $h$ are indicated in the legend. } 
\label{fig9 }

\end{figure} 

\section{Summary} 

In this paper, we use exact matrix diagonalization
to explore the many-body localization properties of the
one-dimensional Ising spin-1 chains. We initially explore the properties of the model's eigenstates by examining the excited-state fidelity $ \mathsf{E}[F] $. We investigate two forms of quasi-disorder and compare them with random disordered systems. The results indicate that both random disorder and quasi-disorder can induce the MBL phase transition in the Ising spin-1 chains. However, under the same conditions, the critical points $h_c$ are different in different disordered systems. This implies that the critical point of the MBL phase transition can be tuned by adjusting the form of disorder of the system. By examining the inflection points of fidelity curves in excited states, we can approximately identify the critical points of the MBL phase transition. We also compared the results for the Ising spin-1 chains and the spin-1/2 chains. Perhaps because spin-1 systems have more degrees of freedom compared to spin-1/2 systems, so we found that the MBL phase transition points in both spin-1 chains are larger than the MBL phase transition points in spin-1/2 chains. The results also show that the critical point of the MBL phase transition is also related to the system size. In the Ising spin-1 and spin-1/2 models, the larger the system size, the larger the critical point of the MBL phase transition. We added the next-nearest-neighbor(N-N) and next-next-nearest-neighbor(N-N-N)interactions. After adding the interactions, the critical points of the phase transition all increase in the Ising model. As the interaction increasing, the system needs stronger disorder to occur MBL phase transition, so the critical points increase. Next, we delve into the dynamical properties of the Ising spin-1 chains with the (N-N) coupling term by the dynamical behavior of diagonal entropy (DE), local magnetization and the time evolution of fidelity. Then we can further prove the occurrence of MBL phase transition in this model and distinguish the ergodic phase (thermal phase) and the many-body localized phase well. The results show that the disordered Ising spin-1 chains with the (N-N) coupling term system can protect its initial information well in the localized phase if the disorder strength of the system is large enough. We also added periodic driving to the Ising spin-1 and spin-1/2 models to study the effect of periodic driving on many-body localization. By observing the fidelity of excited states, it can be seen that, periodic driving will cause the Ising spin-1 and spin-1/2 models to occur the phase transition between the ergodic phase and the localized phase. And the spin-1/2 chains can be driven more stably from the ergodic phase to the localized phase.

\section{Acknowledgments}

This work was supported by the Plan for Scientific and Technological Development of Jilin Province (No. 20230101018JC), by the NSF of China (Grant No. 62175233), by the NSF of China (Grant No. 62375259), and by the Plan for Scientific and Technological Development of Jilin Province (No. 20220101111JC).

We declare that we have no financial and personal relationship with other people or organizations that can inappropriately influence our work, there in no professional or other personal interest of any nature or kind in any product, service or company that could be construed as influencing the position presented in, or the review of, the manuscript entitled.

The data that support the findings of this study are available from the corresponding author, T.T H, upon reasonable request.

\end{document}